# Polymer Thin Films as Universal Substrates for Extreme Ultraviolet Absorption Spectroscopy of Molecular Transition Metal Complexes

**Yusef Shari'ati[a] and Josh Vura-Weis[a]\***

[a]Department of Chemistry, University of Illinois at Urbana-Champaign, Urbana, IL, USA

Correspondence email: vuraweis@illinois.edu

**Synopsis** A polymer thin film deposition method is shown to produce high-quality substrates suitable for both static and time-resolved M-edge XANES spectroscopy.

**Abstract** Polystyrene and polyvinyl chloride thin films are explored as sample supports for extreme ultraviolet (XUV) spectroscopy of molecular transition metal complexes. Thin polymer films prepared by slip-coating are flat, smooth, and transmit much more XUV light than silicon nitride windows. Analytes can be directly cast onto the polymer surface, or codeposited within it. The M-edge x-ray absorption near-edge structure (XANES) spectra (40-90 eV) of eight archetypal transition metal complexes (M=Mn, Fe, Co, Ni) are presented to demonstrate the versatility of this method. The films are suitable for pump/probe transient absorption spectroscopy, as shown by the excited-state spectra of Fe(bpy)$_3$$^{2+}$ in two different polymer supports.




## 1. Introduction

X-ray absorbance near-edge structure (XANES) spectroscopy is a powerful tool for measuring the electronic structure of transition metal complexes. 3$d$ metals are most commonly probed using synchrotron sources at the K and L$_{2,3}$ edges, corresponding to 1$s$→valence and 2$p$→valence transitions. M-edge XANES, which probes 3$p$→valence transitions with energies between 30 and 100 eV, is much rarer due to the short penetration depth of extreme ultraviolet (XUV) photons. Over the past few years, however, the maturation of femtosecond tabletop XUV sources has renewed interest in this spectral range. The large overlap between 3$p$ and 3$d$ orbitals contributes to highly featureful and informative spectra, from which details on the oxidation state, spin state, and ligand field of the metal-containing compound can be extracted (Groot & Kotani, 2008; Zhang *et al.*, 2016).







M-edge XANES has been successfully used to measure excited-state dynamics in transition metal oxides (Vura-Weis *et al.*, 2013; Jiang *et al.*, 2014; Cirri *et al.*, 2017; Schiffmann *et al.*, 2020), and coordination complexes (Chatterley *et al.*, 2016; Ryland *et al.*, 2018, 2019; Ash *et al.*, 2019; Zhang *et al.*, 2019). XUV transient absorption spectroscopy has also been performed on semiconductors such as Si, Ge, and organohalide lead perovskites (Cushing *et al.*, 2019; Lin *et al.*, 2017; Principi *et al.*, 2018; Cushing *et al.*, 2020).

Further adoption of M-edge XANES for molecular samples is hindered by the difficulty in sample preparation. In L- and K-edge absorption spectroscopy, samples are commonly prepared as a fine powder spread by hand onto layers of Kapton (polyimide) tape, or simply pressed into 1 mm sample mounts with windows of the same material (Bunker, 2010). Solution-phase spectroscopy is also possible using fluid jets or flow cells (Wilson *et al.*, 2001; Sham *et al.*, 1989; Smith & Saykally, 2017). Unlike soft and hard X-rays, however, the attenuation length of XUV light is on the order of only tens of nanometers (Henke *et al.*, 1993); sample manipulation at this scale is challenging and the aforementioned preparation procedures do not easily apply. To date, XUV absorption spectroscopy has been limited to gas-phase molecules or those that can be deposited onto a suitable ultrathin substrate using gas-phase methods such as thermal evaporation.

Silicon nitride ("SiN", nominally $Si_3N_4$) is one such substrate, a material traditionally used for transmission windows in XAS (Borja *et al.*, 2016; Dwyer & Harb, 2017; Törmä *et al.*, 2013). SiN is hard, inert, and can be fabricated with subnanometer surface roughness. SiN is not, however, particularly XUV transmissive – even a free-standing 100 nm membrane attenuates 60 eV light by 75%. At these thicknesses the substrates are extremely fragile and difficult to work with. Moreover, most solvents do not wet the SiN surface, which greatly hinders the casting of analyte films through techniques such as spincoating (Norrman *et al.*, 2005). Even when spincoating does not fail, the resultant film may be deposited unevenly due to flexural standing wave patterns that spontaneously arise in the rotating SiN membrane (Advani, 1967). This film inhomogeneity interacts problematically with probe beam spatial chirp and introduces artifacts into the spectrum (Lin *et al.*, 2016). Physical vapor deposition processes sidestep the issues with deposition from solution, but are not suitable for delicate molecules. Thermal evaporation, for example, fails with compounds that are temperature-sensitive or have counterions; these decompose before subliming. This difficulty has inspired great creativity in sample preparation, such as the deposition of cobalt oxide nanoparticles from cryogenic He nanodroplets (Schiffmann *et al.*, 2020), or the use of reflection-absorption spectroscopy from Fe(III) complexes dissolved in glycerol (Lin *et al.*, 2019).

In this work we introduce new methods of sample preparation that enable XUV absorption spectroscopy to be performed on a broad range of transition metal complexes that were previously inaccessible. We show that polymer films are an attractive alternative substrate to SiN. Compared to SiN, polymer thin films are flexible, are wetted by many solvents, and in the process described here





are simple to fabricate without specialized equipment. Furthermore, the XUV transmission of the polymers explored here, polystyrene (PS) and polyvinylchloride (PVC), is respectively ~60% and ~100% greater than SiN for the same thickness.

Analytes may be cast directly onto a polymer thin film surface or may be codeposited from solution. The latter dispersal method embodies a "solid solution" in which analyte molecules are well-separated from one another and is especially useful for compounds that show large electronic changes or fluorescence quenching upon aggregation in the solid state.

## 2. Results and Discussion

**M-edge XANES**. The XUV probe is generated via high-harmonic generation (HHG) in an instrument described previously (Zhang *et al.*, 2016). Briefly, a Ti:sapphire laser produces 4 mJ pulses of 800 nm light at 1 kHz repetition rate with a pulse width of 35 fs FWHM. The IR laser pulses are focused into a semi-infinite gas cell filled with either neon or argon. Figure S1 shows the XUV continuum created in the HHG process. The continuum has intensity in the range 40-90 eV, with the flux maximum at an energy that depends on the gas used. The 525 nm pump for transient experiments is produced by diverting a 0.7 mJ portion of the Ti:sapphire beam to a noncollinear optical parametric amplifier (TOPAS White). A stream of low-pressure nitrogen gas was passed across samples to avoid pump-induced heating.

### 2.1. Polymer Thin Films as Substrates and Matrices

**Fabrication.** While polymer films may be prepared in many ways, this study focuses on film fabrication by "slip-coating" (Figure 1, steps 1-3), a procedure in which liquid solution is drawn out from between sliding glass plates. Slip-coating – like dip-coating, doctor-blading, or flow-coating (Stafford *et al.*, 2006) – is a meniscus-guided deposition technique (Gu *et al.*, 2018). The solution to be cast is loaded between two horizontal parallel plates, with the top plate freely supported by the liquid beneath. Capillary forces constrain the liquid and cause it to evenly coat the plates, while the gap height is maintained approximately constant by the incompressible volume of liquid. The top plate is slipped off manually and frictional drag forces draw out the solution, leaving a wet film on the surface. Evaporation yields the dry thin film. The freestanding film is obtained by delamination from the plate surface, using adhesive tape (Figure 1, steps 5-6) or by slowly immersing the plate into water at an angle of 45 degrees. Slip-coating is simple, fast, inexpensive, and yields films of good quality for XUV transmission absorbance spectroscopy. The film thickness is readily determined by fitting the visible-light interference pattern (Huibers & Shah, 1997), as detailed in the supporting information.





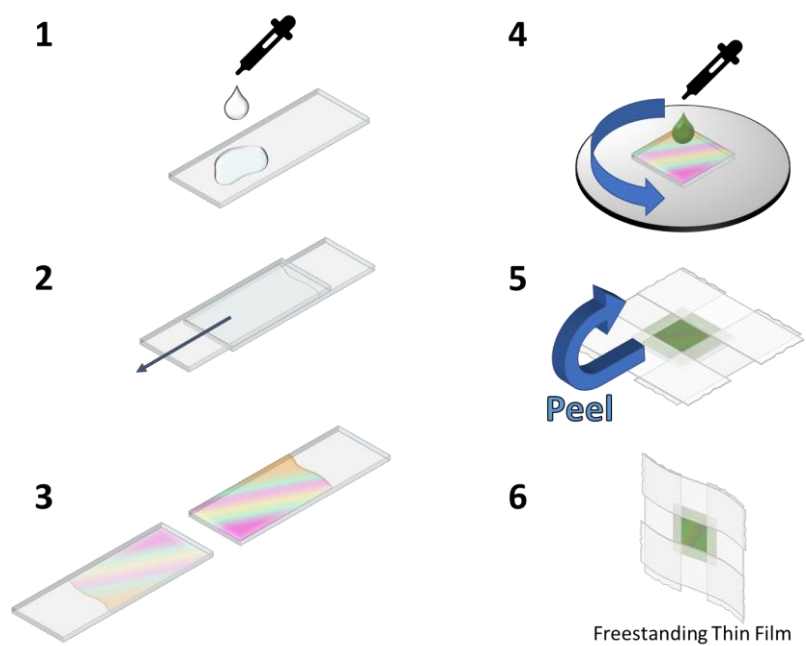

**Figure 1** Schematic of the slip-coating process and thin film delamination by adhesive tape.

## 2.2. Characterization of Polymer Films

The ideal sample substrate has a large XUV transmissivity, is smooth and homogeneous over the length scales of both the XUV probe and the overall sample, and accommodates a wide variety of analytes either within its matrix or upon its surface. This section evaluates the degree to which polymer films fulfil these criteria.

### 2.2.1. XUV Characterization

Figure 2A shows the XUV absorbance profile of a typical PS film. The thickness varies only slightly over the 9 mm$^2$ area, with a standard deviation of 2.2%. The film thickness is therefore essentially constant on the scale of the ~75 µm FWHM XUV beam – an important criterion for mitigating spectral artifacts which arise from sample and probe beam spatial inhomogeneities (Lin *et al.*, 2016). Unlike films prepared from spincoating on flexible substrates, there is no evidence of standing wave patterns in the thickness profile.

The XUV absorbance spectrum was acquired and compared to that of 100 nm SiN membranes. Figure 2B shows the XUV spectrum of these materials, which in the energy range observed comprises only non-resonant absorption due to photoionization of valence electrons. This photoionization is well approximated by a power law and the absorbance is found to be directly proportional to thickness, as shown in Figure 2CC. While the polymers absorb significantly less XUV radiation than does SiN per unit thickness (~40% less for PS and ~50% less for PVC at 60 eV), calculations based on atomic scattering values (shown in Figure 2D) underestimate polymer absorbance, and overestimate SiN absorbance (CXRO Database; Henke *et al.*, 1993). In the case of the polymers, this discrepancy might





be accounted for by a different thin film density from that of the bulk polymer, or by increased photoionization cross section of the polymer molecular orbitals compared to isolated atoms (Vignaud *et al.*, 2014). In contrast to the simulation of $Si_3N_4$, the material used, "SiN," is actually substoichiometric in nitrogen. Indeed, a formulation of $Si_3N_{3.4}$ better fits the data.

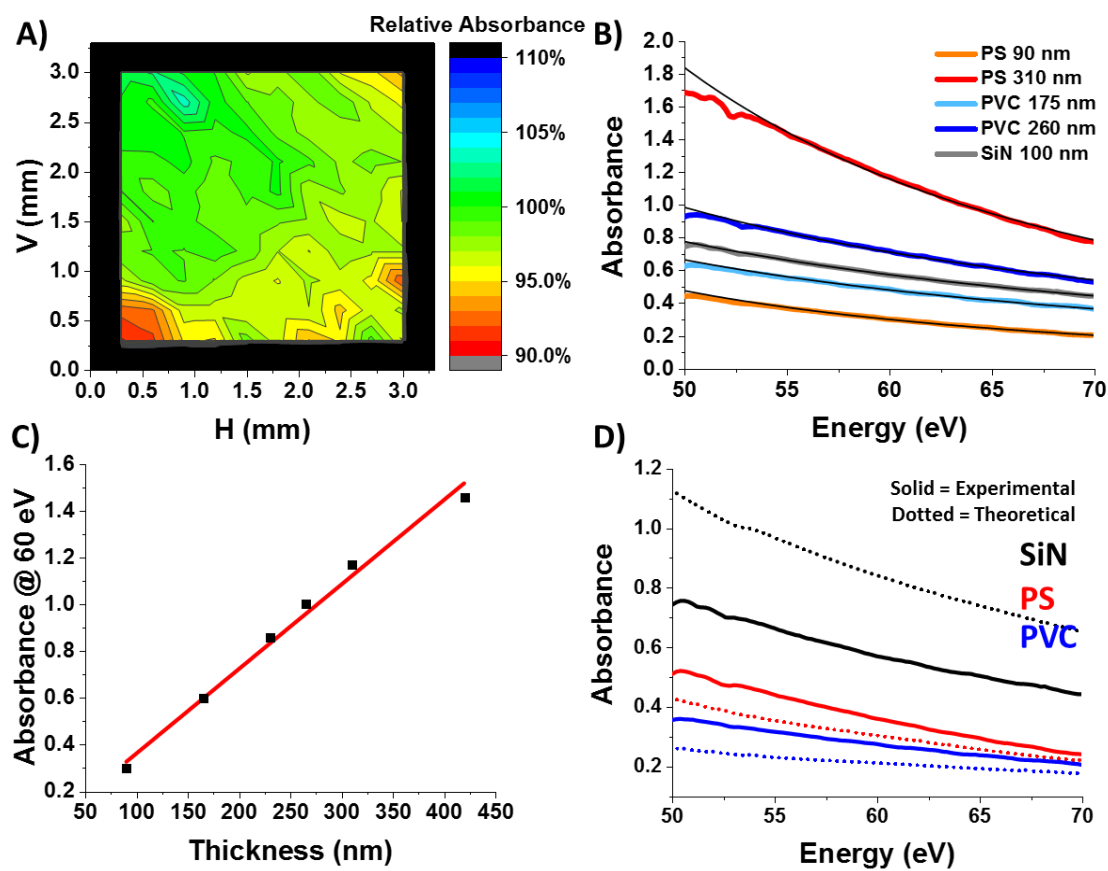

**Figure 2** (A) A 225 nm polystyrene thin film mounted on an empty Si frame. The XUV absorbance was sampled on a 200 μm interval grid to build up an image. Contour lines indicate 1% changes in relative absorbance. (B) Polymer and SiN film XUV spectra. Black lines are power law fits to the data. (C) The XUV absorbance of PS as a function of thickness. The solid line is a linear fit to the data with *y*-intercept = 0 and a slope of $(3.61\pm0.08)\times10^{-3}\,nm^{-1}$. (D) XUV spectra of 100 nm thick samples, as predicted from CXRO data (dotted lines) vs. spectra constructed from the experimentally determined absorption coefficients (solid lines).

### 2.2.2. Polymer Films as Substrate

PS and PVC films present a surface more easily wetted by organic solvents than SiN. Before the film is delaminated, various analyte compounds may be cast upon it by e.g. spin-, slip-, or drop-casting. However, exposure to solvents that dissolve or swell the polymer (such as dichloromethane or tetrahydrofuran) mars the surface and/or prevents the film from delaminating. This can be avoided by appropriate solvent choice and by reducing the time the solvent is in contact with the polymer.





Typically, we accomplish this by spin-coating (Figure 1, step 4) with immediate application of a heat gun to quickly remove solvent, which also limits crystallite size and results in a smoother film. We found that a 4:2:1 mixture of methanol:isopropanol:butanol is an effective solvent system for many analytes; it also evaporates quickly and efficiently wets the polymer surfaces, yet is slow to mar them.

After a second layer is cast upon the polymer base layer, the resulting bilayered film can be delaminated in one piece. PVC films readily delaminate from glass with adhesive tape, while water flotation delaminates both PVC and PS (Figure 1, steps 4 or 5). In some cases, PS films may also be removed with adhesive tape though they generally adhere to the glass more strongly. All PVC bilayer films prepared in this study were removed with adhesive tape.

### 2.2.3. Polymer Films as Matrix

Samples may be prepared by codeposition into a single-layer film, in which the polymer acts as a bulk matrix supporting the analyte. With high analyte loading, the resultant films are no longer removable with adhesive tape and are prone to tearing. Such films must be delaminated by water flotation. If water-soluble, some analyte inevitably leaches out of films when prepared in this way. Very water-sensitive analytes are better prepared on PVC films, which can be delaminated mechanically.

### 2.3. Representative Molecules

The versatility of this sample preparation method is shown using the eight coordination complexes shown in Figure 3, representing several archetypes of molecules that can now be studied easily using M-edge XANES.





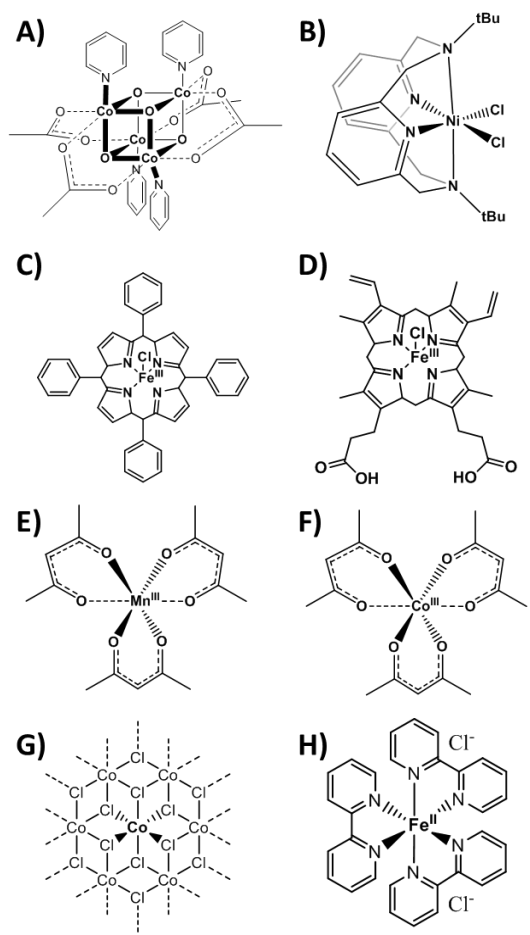

**Figure 3** Chemical structures of (A) cobalt cubane [$Co^{III}_4O_4$], (B) ($tBuN_4$)$Ni^{II}Cl_2$, (C) $Fe^{III}$TPPCl, (D) hemin, (E) $Mn^{II}(acac)_3$, (F) $Co^{III}(acac)_3$, (G) $Co^{II}Cl_2$, and (H) $Fe^{II}(bpy)_3Cl_2$.

### 2.3.1. Ground State XUV Spectroscopy

The M-edge XANES spectrum of each sample was collected and all are shown in Figure 4. Each spectrum shows a resonant absorption edge whose position is primarily determined by element identity and oxidation state. Multiplet features on top of the edge are shaped by analyte oxidation state, spin state, and coordination geometry (Zhang *et al.*, 2016). Spectra have been baselined by subtraction of a power law fit to the pre-edge region, corresponding to the non-resonant photoionization of substrate and ligand atoms. Following the main feature, metal $3p$ photoionization to the continuum contributes to the non-zero cross-section and diminishes approximately as a power law thereafter.

**A.** The molecular cobalt oxide cluster [$Co^{III}_4O_4$](OAc)$_4$(py)$_4$ (OAc = acetate, py = pyridine), or "**cubane**" (Figure 3A) has garnered much interest since its isolation (Beattie *et al.*, 1998) as a model for the cobalt-phosphate (CoPi) system of Nocera (Kanan & Nocera, 2008) as well as a potential water oxidation catalyst in its own right (Smith *et al.*, 2015; Nguyen *et al.*, 2015; Ullman *et al.*, 2014; Nguyen *et al.*, 2017). One of the few complexes capable of supporting a stable Co(IV) center





(Mcalpin *et al.*, 2011), cubane and its oxidized analogues have been the subject of prior X-ray absorption studies, including Co $1s3p$ (K$\beta$) RIXS, which probes the same final state as M-edge XANES (Brodsky *et al.*, 2017; Hadt *et al.*, 2016). In those studies, samples were prepared as either the solid powder in a 1 mm cell with Kapton windows, or as 2 mM solutions in acetonitrile contained within a 3D-printed spectroelectrochemical cell.

Cubane decomposes at temperatures far too low (ca. 120 °C) for sample preparation by thermal evaporation but is easily prepared with polymer films. We now report in Figures 4A1-A2 the ground-state M-edge XANES spectra of cubane samples, both incorporated into PS films and deposited upon PVC. In either case the low-spin cobalt atoms of cubane exhibit a main peak at 64.3 eV and a second at 73 eV. The position and intensity of the features are very similar between sample preparation methods, indicating an insensitivity to the sample environment and no interference from the polymer substrate.

**B.** The complex **($^{tBu}$N4)Ni$^{II}$Cl$_2$** ($^{tBu}$N4 = N,N′-ditert-butyl-2,11-diaza[3.3](2,6)pyridinophane) in Figure 3B serves as an example of an organic-soluble nickel-containing compound, and is a convenient starting point for the formation of Ni$^{I}$ and Ni$^{III}$ catalysts relevant to Kumada and Negishi cross-coupling reactions (Khusnutdinova *et al.*, 2013; Zheng *et al.*, 2014). As shown in Figure 4B, the M-edge XANES spectrum of ($^{tBu}$N4)Ni$^{II}$Cl$_2$ has two prominent peaks at 66.2 and 68.9 eV, consistent with prior reports of Ni$^{II}$ compounds with triplet ground states (Wang *et al.*, 2013; Cirri *et al.*, 2017).

**C-D.** Porphyrins have been studied for light harvesting (Imahori, 2004), phototherapy (Josefsen & Boyle, 2008), and as catalysts for diverse reactions such as oxygen or hydrogen evolution (Zhang *et al.*, 2017). The two porphyrin compounds examined here, iron(III) tetraphenyl porphyrin chloride (**Fe$^{III}$TPPCl**) and iron(III) protoporphyrin IX chloride (**hemin**), are shown in Figures 3C-D. In a previous study from our lab, Fe$^{III}$TPPCl samples were prepared by thermal evaporation and the ultrafast relaxation dynamics were investigated by transient M-edge XANES (Ryland *et al.*, 2018), but the carboxylic acid groups on hemin precludes its sublimation at reasonable temperatures.

The low solubility of Fe$^{III}$TPPCl limits the concentration of material, and hence the signal strength, that can be achieved in PS matrix. Deposition upon PVC was also problematic as dichloromethane (DCM) – one of the best solvents for porphyrins – is not suitable for constructing bilayers in this way. Even very short exposure to DCM mars the smooth polymer surface and prevents delamination from glass. However, it was found that a thin Fe$^{III}$TPPCl film could be spin-coated onto glass from DCM and this neat Fe$^{III}$TPPCl film easily delaminates and floats when slowly immersed into water. This process is like that of polymer delamination and may be generally applicable to hydrophobic glass coatings (Khodaparast *et al.*, 2017). The neat Fe$^{III}$TPPCl film, which is estimated to be <100 nm by its grey-to-golden reflection, is exceedingly brittle and cannot be lifted from the water's surface without destruction. However, it can be successfully picked up upon a thin film





support brought up from below. Multiple layers of sample can be built up by repeating the process as desired. This method was employed to produce the PVC-supported sample of Figure 4C.

Hemin, in contrast, did not require this layer-by-layer process: its carboxylic acid side-groups enable its facile dissolution in alkaline solutions. Samples were prepared by spincoating upon PVC from a mixed alcohol solution containing a small amount of triethylamine.

As shown in Figure 4C, the main peak of $Fe^{III}$TPPCl appears at 57 eV, with a smaller pre-edge feature at 53.8 eV. The position of the main peak in hemin is identical to $Fe^{III}$TPPCl, however its trailing edge diminishes more slowly and widens the feature. The hemin pre-peak (Figure 4D), if it exists, is not resolved.

**E-F.** M(acetylacetonate)$_3$ complexes are classic coordination compounds and lend themselves well to fundamental investigations into electronic structure (Diaz-Acosta *et al.*, 2001, 2003; Carlotto *et al.*, 2017) and validation of spectroscopic techniques (Kubin, Kern *et al.*, 2018; Kubin, Guo *et al.*, 2018; Zhang *et al.*, 2016). The low-spin **Co$^{III}$(acac)$_3$** served as a useful reference in a study showcasing the M-edge XANES spectrum of several cobalt compounds. Samples were prepared by thermal evaporation onto SiN membranes (Zhang *et al.*, 2016). This and the high-spin **Mn$^{III}$(acac)$_3$** are similarly used here, with samples prepared by codeposition from PS solution. Figure 4E shows the M-edge XANES spectrum of Mn$^{III}$(acac)$_3$, whose main feature centers at 52.6 eV. A slight shoulder cleaves from this main peak at 49.0 eV. The spectrum of the cobalt analogue is shown in Figure 4F, and displays a three-peaked structure, with peaks at 64.0, 67, and 74 eV. In comparison with the similarly low-spin $d^6$ cobalt cubane spectra of Figures 4A1-A2, the peaks of Co$^{III}$(acac)$_3$ are at similar positions though are slightly sharper and differ in relative intensity, likely due to the more rigidly octahedral symmetry of the acetylacetonate complex.

**G. Co$^{II}$Cl$_2$** is a high-spin polymeric ionic compound with the $Co^{2+}$ ions assuming octahedral geometry (Figure 3G). Aside from X-ray studies motivated by fundamental interest in its ground-state electronic configuration (Kikas *et al.*, 1999; Wang *et al.*, 2017), Co$^{II}$Cl$_2$ is useful as a precursor in the production of cobalt oxide and cobalt metal thin films (Väyrynen *et al.*, 2018). The M-edge XANES spectrum of Co$^{II}$Cl$_2$ is shown in Figure 4G, displaying two large features at 61.4 and 63.8 eV, and a smaller at 58.6 eV.

**H. Fe$^{II}$(bpy)$_3$Cl$_2$** (bpy = 2,2'-bipyridine, Figure 3H) and similar Fe polypyridyl complexes are the subject of intense scrutiny due to their ultrafast intersystem crossing rates. Although Ruthenium polypyridyl complexes have metal-to-ligand charge-transfer (MLCT) states with lifetimes on the order of hundreds of nanoseconds to a microsecond (Juris *et al.*, 1988), their iron congeners relax in less than 200 fs to low-energy triplet and quintet metal-centered states (Auböck & Chergui, 2015). The former are excellent chromophores in dye-sensitized solar cells and in photoredox chemistries but the latter compounds, attractively earth-abundant and inexpensive, are inefficient due to these short





lifetimes (Ardo & Meyer, 2009; Wenger, 2019; McCusker, 2019). This discrepancy has driven research into better understanding the excited-state surfaces that drive these dynamics (Miaja-Avila *et al.*, 2016; Zhang *et al.*, 2019; Auböck & Chergui, 2015). The $Fe^{II}(bpy)_3Cl_2$ samples were here prepared by spincoating from solution onto PVC. The ground-state M-edge XANES spectrum (Figure 4H) displays three main peaks and a shoulder at 58.4, 61.6, 67.4, and 55.0 eV respectively.

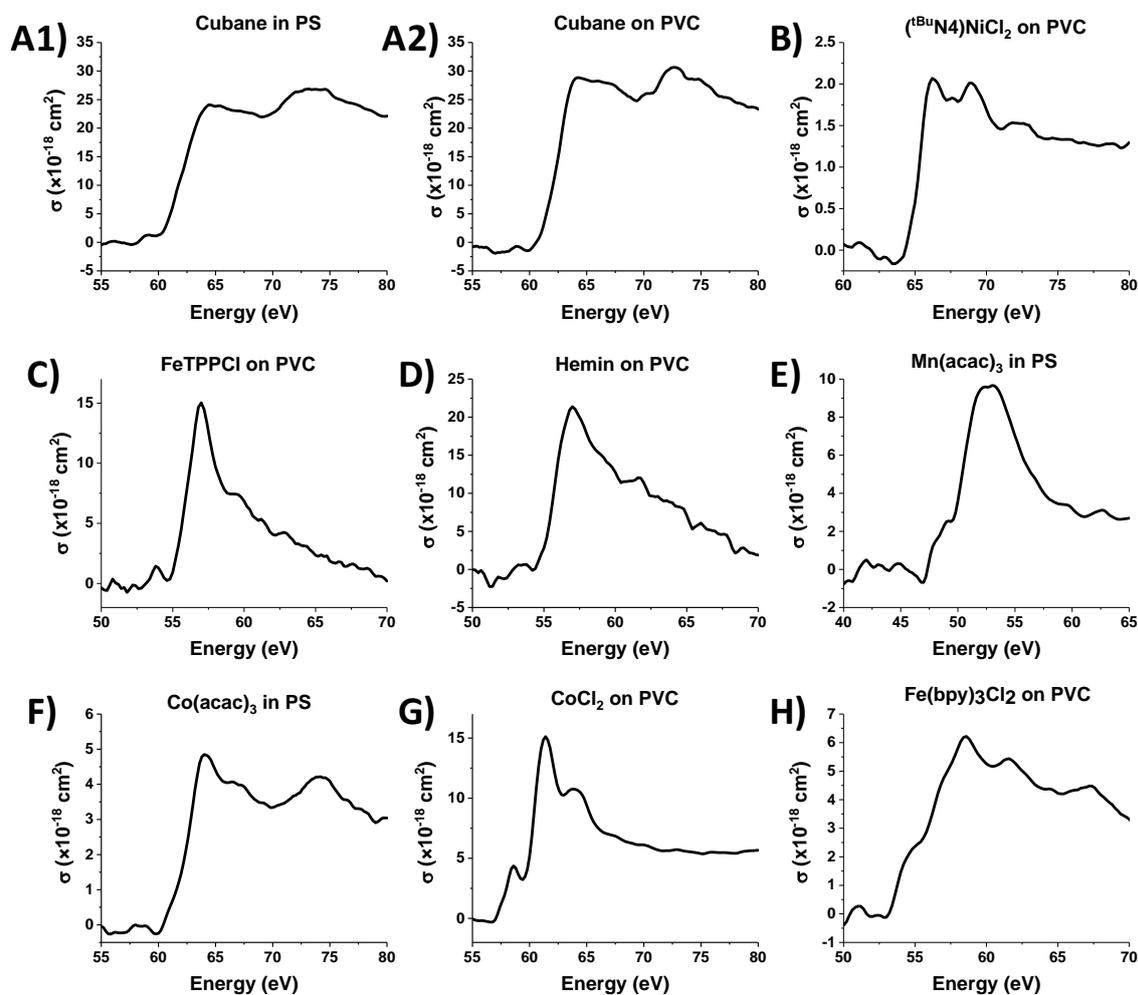

**Figure 4** XUV spectra obtained from analytes deposited on the surface of PVC, or codeposited in PS.

### 2.3.2. Transient XUV Spectroscopy

Efforts to improve iron(II) polypyridyl complexes have focused on altering the problematic intermediate states on the relaxation pathway through rational ligand design (Chábera *et al.*, 2018; Wenger, 2019; McCusker, 2019). However, directly observing and characterizing these states requires a technique with ultrafast time-resolution and spin sensitivity. We recently used M-edge XANES to identify an intermediate $^3T$ state and coherent oscillations on the $^5T_{2g}$ surface in $Fe^{II}(phen)_3(SCN)_2$ (phen = *o*-phenanthroline) (Zhang *et al.*, 2019). Unlike the fortuitously sublimable phenanthroline





compound, thin films of other iron(II) polypyridyl compounds, including the oft-studied prototypical spin-crossover compound $Fe^{II}(bpy)_3Cl_2$, are not so easily prepared.

The sample preparation methods developed here enabled collection of the transient M-edge XANES spectra of two $Fe^{II}(bpy)_3^{2+}$ compounds: $Fe^{II}(bpy)_3Cl_2$ cast on PVC and $Fe^{II}(bpy)_3(PF_6)_2$ codeposited in PS. Samples were pumped into the MLCT band at 525 nm and the difference spectra collected at delay times when the $^5T$ state is fully populated, between 1.0 and 2.0 ps (Auböck & Chergui, 2015). Figure 5 shows the difference spectra of these two $Fe^{II}(bpy)_3^{2+}$ samples in comparison to the previously-published $^5T_{2g}$ difference spectrum of $Fe^{II}(phen)_3(SCN)_2$. The spectra exhibit a positive excited state absorption signal near 57.1 eV with a shoulder at 55.3 eV that is more sharply defined in the bipyridine complexes. Each spectrum also displays a ground-state bleach near 67.5 eV. The successful acquisition of these spectra underscores the versatility of polymer films as well as the aptitude of M-edge XANES spectroscopy towards the determination of excited state electronic structure in metal complexes. The way is now made clear towards the future measurement of further spin crossover compounds.

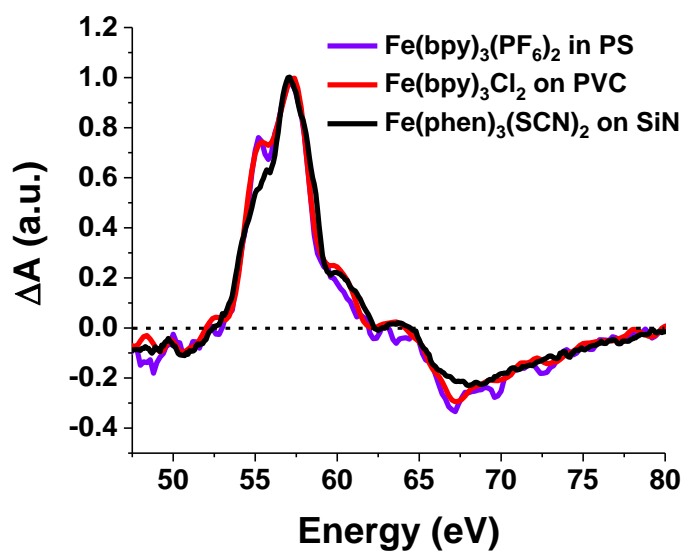

**Figure 5** The normalized excited state difference spectra of iron polypyridyl compounds. Red: $Fe(bpy)_3Cl_2$ on PVC, time-averaged between 1.0 and 2.0 ps. Purple: $Fe^{II}(bpy)_3(PF_6)_2$ in PS, time-averaged between 1.0 and 2.0 ps. Black: spectral component of global fit to $Fe^{II}(phen)_3(SCN)_2$ data corresponding to the $^5T_{2g}$ state.

## 3. Conclusion

The growing development of HHG sources of XUV radiation permits core-level spectroscopy to be performed using convenient, in-lab instruments at femtosecond to attosecond timescales (Geneaux *et al.*, 2019; Zhang *et al.*, 2016). These advantages are balanced by the requirements that samples be ultrathin and smooth. Limited sample preparation protocols have held back development in this area





and M-edge XANES spectroscopy has so far been restricted to those materials evaporable onto hard substrates such as $Si_3N_4$ or otherwise amenable to gas-phase experiments. We have now demonstrated a new technique for solution-processable sample deposition, enabling acquisition of M-edge spectra of many previously inaccessible compounds, shown here in the example of eight model compounds. The polystyrene and polyvinyl chloride polymer substrates used here are easily fabricated in a simple slip-coating technique. They are flexible, highly XUV transmissive, and have more favorable wetting properties than silicon nitride. The polymers accommodate a wide variety of analytes, either directly cast upon the surface or dissolved within the bulk of the polymer thin film. Such samples are homogeneous, ultrathin, and smooth, mitigating harmonic spectral artifacts and enabling the acquisition of high-quality ground and excited state XUV spectra. This technique significantly broadens the types of molecular complexes that can be studied with M-edge XANES.

**Acknowledgements**    This material is based upon work supported by the U.S. Department of Energy, Office of Science, Office of Basic Energy Sciences under Award Number DE-SC0018904. This material is based upon work supported by the National Science Foundation Graduate Research Fellowship under Grant No. DGE-1746047. The XUV instrument was built with partial funding from the Air Force Office of Scientific Research under AFOSR Awards No. FA9550-14-1-0314 and FA9550-18-1-0293.

**References**

Advani, S. H. (1967). *Int. J. Mech. Sci.* **9**, 307–313.

Ardo, S. & Meyer, G. J. (2009). *Chem. Soc. Rev.* **38**, 115–164.

Ash, R., Zhang, K. & Vura-Weis, J. (2019). *J. Chem. Phys.* **151**, 104201.

Auböck, G. & Chergui, M. (2015). *Nat. Chem.* **7**, 629–633.

Beattie, J. K., Hambley, T. W., Klepetko, J. A., Masters, A. F. & Turner, P. (1998). *Polyhedron*. **17**, 1343–1354.

Borja, L. J., Zürch, M., Pemmaraju, C. D., Schultze, M., Ramasesha, K., Gandman, A., Prell, J. S., Prendergast, D., Neumark, D. M. & Leone, S. R. (2016). *J. Opt. Soc. Am. B*. **33**, C57.

Brodsky, C. N., Hadt, R. G., Hayes, D., Reinhart, B. J., Li, N., Chen, L. X. & Nocera, D. G. (2017). *Proc. Natl. Acad. Sci.* **114**, 3855–3860.

Bunker, G. (2010). Introduction to XAFS : a practical guide to X-ray absorption fine structure spectroscopy Cambridge: Cambridge University Press.

Carlotto, S., Floreano, L., Cossaro, A., Dominguez, M., Rancan, M., Sambi, M. & Casarin, M. (2017). *Phys. Chem. Chem. Phys.* **19**, 24840–24854.





Chábera, P., Kjaer, K. S., Prakash, O., Honarfar, A., Liu, Y., Fredin, L. A., Harlang, T. C. B., Lidin, S., Uhlig, J., Sundström, V., Lomoth, R., Persson, P. & Wärnmark, K. (2018). *J. Phys. Chem. Lett.* **9**, 459–463.

Chatterley, A. S., Lackner, F., Pemmaraju, C. D., Neumark, D. M., Leone, S. R. & Gessner, O. (2016). *J. Phys. Chem. A.* **120**, 9509–9518.

Cirri, A., Husek, J., Biswas, S. & Baker, L. R. (2017). *J. Phys. Chem. C.* **121**, 15861–15869.

Cushing, S. K., Lee, A., Porter, I. J., Carneiro, L. M., Chang, H. T., Zürch, M. & Leone, S. R. (2019). *J. Phys. Chem. C.* **123**, 3343–3352.

Cushing, S. K., Porter, I. J., de Roulet, B. R., Lee, A., Marsh, B. M., Szoke, S., Vaida, M. E. & Leone, S. R. (2020). *Sci. Adv.* **6**, eaay6650.

Diaz-Acosta, I., Baker, J., Cordes, W. & Pulay, P. (2001). *J. Phys. Chem. A.* **105**, 238–244.

Diaz-Acosta, I., Baker, J., Hinton, J. F. & Pulay, P. (2003). *Spectrochim. Acta - Part A Mol. Biomol. Spectrosc.* **59**, 363–377.

Dwyer, J. R. & Harb, M. (2017). *Appl. Spectrosc.* **71**, 2051–2075.

Geneaux, R., Marroux, H. J. B., Guggenmos, A., Neumark, D. M. & Leone, S. R. (2019). *Philos. Trans. R. Soc. A Math. Phys. Eng. Sci.* **377**,.

Groot, F. de & Kotani, A. (2008). Core Level Spectroscopy of Solids CRC Press.

Gu, X., Shaw, L., Gu, K., Toney, M. F. & Bao, Z. (2018). *Nat. Commun.* **9**,.

Hadt, R. G., Hayes, D., Brodsky, C. N., Ullman, A. M., Casa, D. M., Upton, M. H., Nocera, D. G. & Chen, L. X. (2016). *J. Am. Chem. Soc.* **138**, 11017–11030.

Henke, B. L., Gullikson, E. M. & Davis, J. C. (1993). *At. Data Nucl. Data Tables.* **54**, 181–342.

Huibers, P. D. T. & Shah, D. O. (1997). *Langmuir.* **13**, 5995–5998.

Imahori, H. (2004). *J. Phys. Chem. B.* **108**, 6130–6143.

Jiang, C. M., Baker, L. R., Lucas, J. M., Vura-Weis, J., Alivisatos, A. P. & Leone, S. R. (2014). *J. Phys. Chem. C.* **118**, 22774–22784.

Josefsen, L. B. & Boyle, R. W. (2008). *Met. Based. Drugs.* **2008**,.

Juris, A., V., B., Barigelletti, F., Campagna, S., Belser, P. & von Zelewsky, A. (1988). *Coord. Chem. Rev.* **84**, 85–277.

Kanan, M. W. & Nocera, D. G. (2008). *Science (80-. ).* **321**, 1072–1075.

Khodaparast, S., Boulogne, F., Poulard, C. & Stone, H. A. (2017). *Phys. Rev. Lett.* **119**, 1–5.

Khusnutdinova, J. R., Luo, J., Rath, N. P. & Mirica, L. M. (2013). *Inorg. Chem.* **52**, 3920–3932.

Kikas, A., Ruus, R., Saar, A., Nõmmiste, E., Käämbre, T. & Sundin, S. (1999). *J. Electron Spectros. Relat. Phenomena.* **101–103**, 745–749.

Kubin, M., Guo, M., Ekimova, M., Baker, M. L., Kroll, T., Källman, E., Kern, J., Yachandra, V. K., Yano, J., Nibbering, E. T. J., Lundberg, M. & Wernet, P. (2018). *Inorg. Chem.* **57**, 5449–5462.

Kubin, M., Kern, J., Guo, M., Källman, E., Mitzner, R., Yachandra, V. K., Lundberg, M., Yano, J. & Wernet, P. (2018). *Phys. Chem. Chem. Phys.* **20**, 16817–16827.






Lin, L., Husek, J., Biswas, S., Baumler, S. M., Adel, T., Ng, K. C., Baker, L. R. & Allen, H. C. (2019). *J. Am. Chem. Soc.* **141**, 13525–13535.

Lin, M.-F., Verkamp, M. A., Ryland, E. S., Zhang, K. & Vura-Weis, J. (2016). *J. Opt. Soc. Am. B*. **33**, 1986.

Lin, M. F., Verkamp, M. A., Leveillee, J., Ryland, E. S., Benke, K., Zhang, K., Weninger, C., Shen, X., Li, R., Fritz, D., Bergmann, U., Wang, X., Schleife, A. & Vura-Weis, J. (2017). *J. Phys. Chem. C*. **121**, 27886–27893.

Mcalpin, J. G., Stich, T. a, Ohlin, C. A., Surendranath, Y. & Daniel, G. (2011). *J. Am. Chem. Soc.* **133**, 15444–15452.

McCusker, J. K. (2019). *Science (80-. ).* **363**, 484–488.

Miaja-Avila, L., O'Neil, G. C., Joe, Y. I., Alpert, B. K., Damrauer, N. H., Doriese, W. B., Fatur, S. M., Fowler, J. W., Hilton, G. C., Jimenez, R., Reintsema, C. D., Schmidt, D. R., Silverman, K. L., Swetz, D. S., Tatsuno, H. & Ullom, J. N. (2016). *Phys. Rev. X.* **6**, 1–13.

Nguyen, A. I., Wang, J., Levine, D. S., Ziegler, M. S. & Tilley, T. D. (2017). *Chem. Sci.* **8**, 4274–4284.

Nguyen, A. I., Ziegler, M. S., Oña-Burgos, P., Sturzbecher-Hohne, M., Kim, W., Bellone, D. E. & Tilley, T. D. (2015). *J. Am. Chem. Soc.* **137**, 12865–12872.

Norrman, K., Ghanbari-Siahkali, A. & Larsen, N. B. (2005). *Annu. Reports Prog. Chem. - Sect. C.* **101**, 174–201.

Principi, E., Giangrisostomi, E., Mincigrucci, R., Beye, M., Kurdi, G., Cucini, R., Gessini, A., Bencivenga, F. & Masciovecchio, C. (2018). *Phys. Rev. B*. **97**, 174107.

Ryland, E. S., Lin, M. F., Verkamp, M. A., Zhang, K., Benke, K., Carlson, M. & Vura-Weis, J. (2018). *J. Am. Chem. Soc.* **140**, 4691–4696.

Ryland, E. S., Zhang, K. & Vura-Weis, J. (2019). *J. Phys. Chem. A*. **123**, 5214–5222.

Schiffmann, A., Toulson, B. W., Knez, D., Messner, R., Schnedlitz, M., Lasserus, M., Hofer, F., Ernst, W. E., Gessner, O. & Lackner, F. (2020). *J. Appl. Phys.* **127**, 184303.

Sham, T. K., Yang, B. X., Kirz, J. & Tse, J. S. (1989). *Phys. Rev. A*. **40**, 652–669.

Smith, J. W. & Saykally, R. J. (2017). *Chem. Rev.* **117**, 13909–13934.

Smith, P. F., Hunt, L., Laursen, A. B., Sagar, V., Kaushik, S., Calvinho, K. U. D., Marotta, G., Mosconi, E., De Angelis, F. & Dismukes, G. C. (2015). *J. Am. Chem. Soc.* **137**, 15460–15468.

Stafford, C. M., Roskov, K. E., Epps, T. H. & Fasolka, M. J. (2006). *Rev. Sci. Instrum.* **77**,.

Törmä, P. T., Sipilä, H. J., Mattila, M., Kostamo, P., Kostamo, J., Kostamo, E., Lipsanen, H., Nelms, N., Shortt, B., Bavdaz, M. & Laubis, C. (2013). *IEEE Trans. Nucl. Sci.* **60**, 1311–1314.

Ullman, A. M., Liu, Y., Huynh, M., Bediako, D. K., Wang, H., Anderson, B. L., Powers, D. C., Breen, J. J., Abruña, H. D. & Nocera, D. G. (2014). *J. Am. Chem. Soc.* **136**, 17681–17688.

Väyrynen, K., Hatanpää, T., Mattinen, M., Heikkilä, M., Mizohata, K., Meinander, K., Räisänen, J., Ritala, M. & Leskelä, M. (2018). *Chem. Mater.* **30**, 3499–3507.







Vignaud, G., Chebil, M. S., Bal, J. K., Delorme, N., Beuvier, T., Grohens, Y. & Gibaud, A. (2014). *Langmuir.* **30**, 11599–11608.

Vura-Weis, J., Jiang, C. M., Liu, C., Gao, H., Lucas, J. M., De Groot, F. M. F., Yang, P., Alivisatos, A. P. & Leone, S. R. (2013). *J. Phys. Chem. Lett.* **4**, 3667–3671.

Wang, H., Young, A. T., Guo, J., Cramer, S. P., Friedrich, S., Braun, A. & Gu, W. (2013). *J. Synchrotron Radiat.* **20**, 614–619.

Wang, R. P., Liu, B., Green, R. J., Delgado-Jaime, M. U., Ghiasi, M., Schmitt, T., Van Schooneveld, M. M. & De Groot, F. M. F. (2017). *J. Phys. Chem. C.* **121**, 24919–24928.

Wenger, O. S. (2019). *Chem. - A Eur. J.* **25**, 6043–6052.

Wilson, K. R., Rude, B. S., Catalane, T., Schaller, R. D., Tobin, J. G., Co, D. T. & Saykally, R. J. (2001). *J. Phys. Chem. B.* **105**, 3346–3349.

Zhang, K., Ash, R., Girolami, G. S. & Vura-Weis, J. (2019). *J. Am. Chem. Soc.* **141**, 17180–17188.

Zhang, K., Lin, M. F., Ryland, E. S., Verkamp, M. A., Benke, K., De Groot, F. M. F., Girolami, G. S. & Vura-Weis, J. (2016). *J. Phys. Chem. Lett.* **7**, 3383–3387.

Zhang, W., Lai, W. & Cao, R. (2017). *Chem. Rev.* **117**, 3717–3797.

Zheng, B., Tang, F., Luo, J., Schultz, J. W., Rath, N. P. & Mirica, L. M. (2014). *J. Am. Chem. Soc.* **136**, 6499–6504.






# Supporting information

### S1. Materials & Methods

**Materials.** Polystyrene ($M_w$ = 192,000), $Fe^{III}TPPCl$ and $Co(acac)_3$ were obtained from Sigma-Aldrich, hemin and $Mn(acac)_3$ from Thermo-Fisher Scientific, and polyvinyl chloride ($M_w$ = 275,000) from Scientific Polymer Products, Inc. $CoCl_2 \cdot 6\ H_2O$ was obtained from Acros organics. Cubane (Chakrabarty *et al.*, 2007) and $Fe(bpy)_3Cl_2$ (Jaeger & van Dijk, 1936) were synthesized according to literature procedures, while samples of $(^{tBu}N4)NiCl_2$ were graciously provided by the Mirica group (Khusnutdinova *et al.*, 2013). SiN membrane windows (100 nm, frame 7.5 mm × 7.5 mm, window 3.0 mm × 3.0 mm) were obtained from Silson Ltd. Empty silicon frames (in which the central SiN membrane has broken) were used as sample mounts.

**General Method for Preparation of Polymer Films by Slip-Coating.** Prior to film deposition, 75 mm × 25 mm glass slides (obtained from VWR) are cleaned with isopropanol and dried with a heat gun. Polymer solutions are prepared by sonication of the resin in solvent for at least 30 minutes so that no solids remain, then filtered over silica. A polymer film is deposited by first dispensing ~100 µL of solution on the slide. A second slide is laid upon the first, offset by 1 cm, taking care to introduce no air bubbles. The top slide is then removed in a single swift (ca. 5 cm/s) and steady sliding motion. The two slides, now evenly coated in polymer solution, are left to dry in ambient conditions for 60 s. Thereafter, the slides are further dried with a heat gun for 10 s to ensure complete solvent evaporation. Films are removed from the glass substrate by water flotation, or directly with adhesive tape. The thickness of deposited films is varied by adjusting polymer concentration accordingly, though slide shearing velocity plays a role as well (Stafford *et al.*, 2006).

**Preparation of PVC Films.** A solution consisting of 1.5 wt% PVC in 93.5% THF, 5% cyclohexanone was slip-coated between glass slides. PVC films are delaminated by water flotation or directly with adhesive tape.

**Preparation of PS Films.** Neat polystyrene films were formed from 1.5 wt% polystyrene in dichloromethane (DCM) or 1,2-dichloroethane (DCE) by slip-coating, as above. All PS films were made freestanding by water flotation.

**Water Flotation.** All films prepared may be removed from the glass substrate by gradually lowering the slide into water at an oblique angle. The film floats on the surface of the water. The film is retrieved and mounted onto an empty silicon frame by bringing the latter up from underneath. The sample is gently dried with a heat gun, which also anneals any wrinkles in the freestanding film.

**Spincasting on PVC.** A section of PVC-coated glass (prepared via slip-coating as above) is spun at 1400 rpm and a single drop (~0.05 mL) of analyte solution is made to fall upon the center of rotation.







Film formation is often improved with the immediate application of a heat gun. The resultant two-layer film is delaminated with adhesive tape.

**Specific Film Preparation Conditions**

**[Co$^{III}_4$O$_4$] Cubane in PS:** 118 mg of cubane solution (29 mM in DCM) was combined with 63 mg 3 wt% PS in DCE. The resultant dark green solution was slipcoated.

**[Co$^{III}_4$O$_4$] Cubane on PVC:** 44 mg ($5.16 \times 10^{-5}$ mol) cubane was dissolved in a mixed alcohol solution consisting of 1.20 g MeOH, 0.60 g iPrOH, and 0.30 g BuOH. This ~20 mM solution was spincoated onto PVC-coated glass with the immediate application of a heat gun. The resultant two-layered film was delaminated with adhesive tape.

**($^{tBu}$N4)Ni$^{II}$Cl$_2$:** 6.0 mg ($1.24 \times 10^{-5}$ mol, 482.12 g/mol) was dissolved in a solution of 68 mg MeOH, 40 mg iPrOH, and 14 mg BuOH. The green solution (~80 mM) was filtered and then slipcoated onto a PVC-coated glass slide, with immediate heat gun application after removing the top slide. The film was delaminated with adhesive tape.

**Fe$^{III}$TPPCl:** A 3.6 mM solution (2.1 mg dissolved in 1.05 g DCE), without polymer, was slipcoated. The resultant golden-brown film was delaminated by water flotation and picked up onto a mounted 100 nm PVC film.

**Hemin:** 1.3 mg was dissolved in 80 mg MeOH, 65 mg iPrOH, 19 mg BuOH, and 22 mg triethylamine. The dark brown solution was spincoated onto a PVC-coated glass slide and the bilayer was delaminated with tape.

**Mn$^{III}$(acac)$_3$:** 100 mg of 27 mM compound was mixed with 110 mg of 3.0 wt% PS in DCE and the resulting dark brown solution was slipcoated.

**Co$^{III}$(acac)$_3$:** 70 mg of a 70 mM Co(acac)$_3$ DCE solution was added to 30 mg of 3.0 wt% PS in DCE. This deep green solution was slipcoated.

**Co$^{II}$Cl$_2$:** A blue 5.4 mM acetonitrile solution of CoCl$_2$•6 H$_2$O was spincast upon a mounted PVC membrane with immediate heat gun application to remove solvent. The PVC was then delaminated with adhesive tape.





**Fe$^{II}$(bpy)$_3$Cl$_2$:** 3.6 mg was dissolved in a solution of 70 mg MeOH, 45 mg iPrOH, and 40 mg BuOH. This deep red solution (~30 mM) was filtered, then spuncast with heating onto a PVC-coated glass slide. The film was removed with adhesive tape.

### S2. M-edge XANES

**XUV Probe Generation.** High-harmonic generation (HHG) in a noble gas (Ar or Ne) produces the XUV probe, shown in **Error! Reference source not found.**A. The periodic spikes in the continuum arising every 3.1 eV correspond to the odd harmonics of the 800 nm driving laser. A silicon mirror and 200 nm aluminum filter remove the residual IR, leaving the XUV light to be passed on through to the sample and then dispersed onto a CCD by a diffraction grating. A Zr filter can be exchanged with the Al filter to obtain intensity after the 72.6 eV aluminum absorption edge. The system is kept under high vacuum ($10^{-6}$ torr) to mitigate XUV attenuation. Energy calibration parameters and spectrometer resolution (typically 0.35 eV FWHM) are computed daily from measurements of Fe$_2$O$_3$ (Vura-Weis *et al.*, 2013), NiO (Chiuzăian *et al.*, 2005; Wang *et al.*, 2013), and ionized xenon samples (Andersen *et al.*, 2001).

Pump-probe experiments utilized the output of a noncollinear optical parametric amplifier (TOPAS White) tuned to 525 nm. Power for the pump was 0.62 mJ and 1.65 mJ, for Fe(bpy)$_3$Cl$_2$ and Fe(bpy)$_3$(PF$_6$)$_2$ samples respectively. The pump beam spot size at the sample was 220 µm FWHM and the XUV beam size was 75 µm FWHM as measured by knife-edge scan. The low thermal conductivity of polymer films necessitated that a stream of low-pressure nitrogen gas be passed across samples to avoid pump-induced heating and damage.

**M-Edge XANES Spectra.** XUV absorbance spectra are generated using the base-10 logarithm of the ratio between transmitted counts through the sample and a reference, chosen to be of the same material and similar thickness as the sample substrate. The non-resonant absorbance background due to photoionization of valence electrons is approximated as a power law and removed by subtraction; the power law parameters are obtained from a fit to the pre-edge region. All reported M-edge XANES spectra are baselined in this way, as shown in **Error! Reference source not found.**B.





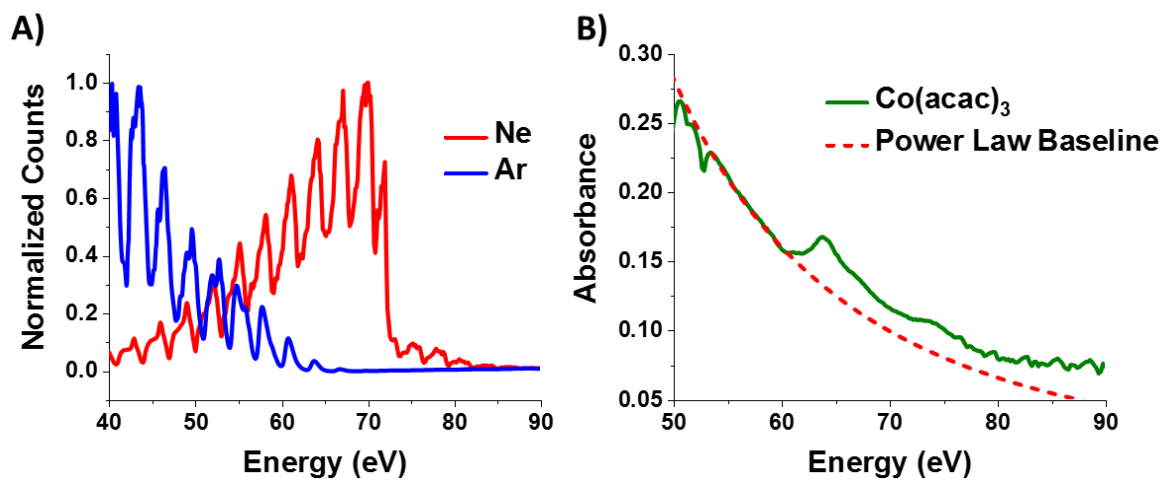

**Figure S1** (A) The typical XUV continuum generated in argon and neon gases. (B) Raw XUV absorption spectrum of Co(acac)$_3$ acquired under neon HHG, shown also with power law fit to non-resonant photoionization background. The deviations from the power law below 55 eV are artifacts due to low photon flux.

**Cross-Section Determination.** The XUV resonant absorption cross-section $\sigma_{XUV}$ was determined by comparison of the XUV and UV-visible spectrum for each sample. The known UV-visible cross-section allows calculation of $\sigma_{XUV}$ via the following

$$\sigma_{XUV} = \sigma_{UV} \frac{A_{XUV}}{A_{UV}} \tag{1}$$

Where, in a particular sample, $A_{XUV}$ is the XUV resonant absorbance and $A_{UV}$ is the absorbance at a prominent wavelength chosen in the analyte's UV-visible spectrum. The cross-section $\sigma_{UV}$ at that wavelength is independently calculated from analyzing separately deposited neat analyte films of known thickness according to equation (2), where $\varepsilon$ is the molar extinction coefficient (L mol$^{-1}$ cm$^{-1}$), MW is the molecular weight, $h$ is the film thickness in nm, and $\rho$ is the film density.

$$\varepsilon = 10^7 (\frac{nm}{cm}) \times \frac{A_{UV} \times MW}{1000 \times h \times \rho} \tag{2}$$

**Power Law Fit Parameters.** The non-resonant XUV absorbance of polymer and SiN films is well approximated by the power law shown in equation (3), where $A$ is absorbance, $\alpha$ is an attenuation coefficient, $d$ is the film thickness, $E$ is energy, and $E_0$ is the reference energy (taken here to be 60 eV). **Error! Reference source not found.** reports these parameters obtained from a least-squares fit to the experimental data for the three substrates described.

$$A = \alpha d \left(\frac{E}{E_0}\right)^{-k} \tag{3}$$





**Table S1**     Power law fit parameters for polymer substrates.

|        | $\alpha$ (nm$^{-1}$) | $k$ |
|--------|---------------------|-----|
| SiN    | $5.69\times10^{-3}$ | 1.65 |
| PS     | $3.61\times10^{-3}$ | 2.53(6) |
| PVC    | $2.76\times10^{-3}$ | 1.79(8) |

**Thin Film Interference.** The effects of interference in thin films result in a wavelength-dependent reflection of incident light which is observable in the UV-visible spectrum. The ratio of transmitted light intensity $I$ to the incident light intensity $I_0$ at a given wavelength $\lambda$, derived from the absorbance $A$, is given by equation (). The parameter $r$ is the Fresnel reflection coefficient, $n$ is the index of refraction of the thin film, and $d$ is the film thickness (Huibers & Shah, 1997).

$$\frac{I}{I_0} = 10^{-A} = \frac{(1-r)^2}{1 + r^2 - 2r\,cos\left(\frac{4\pi n d}{\lambda}\right)} \qquad (4)$$

The Fresnel reflection coefficient at normal incidence is given by equation (), where $n_0$ is the index of refraction of the medium surrounding the thin film, i.e. air ($n_{air} = 1$).

$$r = \left(\frac{n - n_0}{n - n_0}\right)^2 \qquad (5)$$

The index of refraction for PVC ($n_{PVC}$) is nearly a constant 1.53 over the range of the UV-visible spectrometer (190-820 nm). However, the dispersion in PS is significant enough that $n_{PS}$ must be modeled empirically with Cauchy's equation, shown in equation () with literature values for PS (Jones *et al.*, 2013).

$$n_{PS} = 1.5718 + \frac{8412\;nm^2}{\lambda^2} + \frac{2.35 \times 10^8\;nm^4}{\lambda^4} \qquad (6)$$

**Example Film Analysis.** Displayed in **Error! Reference source not found.** is the UV-visible spectrum of a thin film composed of cubane codeposited with PS. The experimental ("sample") spectrum is regarded as a sum of two component spectra: the periodic undulatory spectrum from interference and the absorbance spectrum due to electronic transitions in cubane. PS itself absorbs negligibly in the range displayed. The cubane portion of the total spectrum was obtained from the film prior to delamination (where the interference effect is not observed due to the similar refractive indices of PS and the underlying glass slide). The magnitude of the cross-section $\sigma_{UV}$ of cubane in PS is assumed to be the same as that measured in dichloromethane solution. The refractive index of the sample is assumed to be negligibly changed from that of pure PS. Given the above, the least-squares fit indicates a sample thickness of 505 nm with a cubane concentration of 0.65 mol/L.

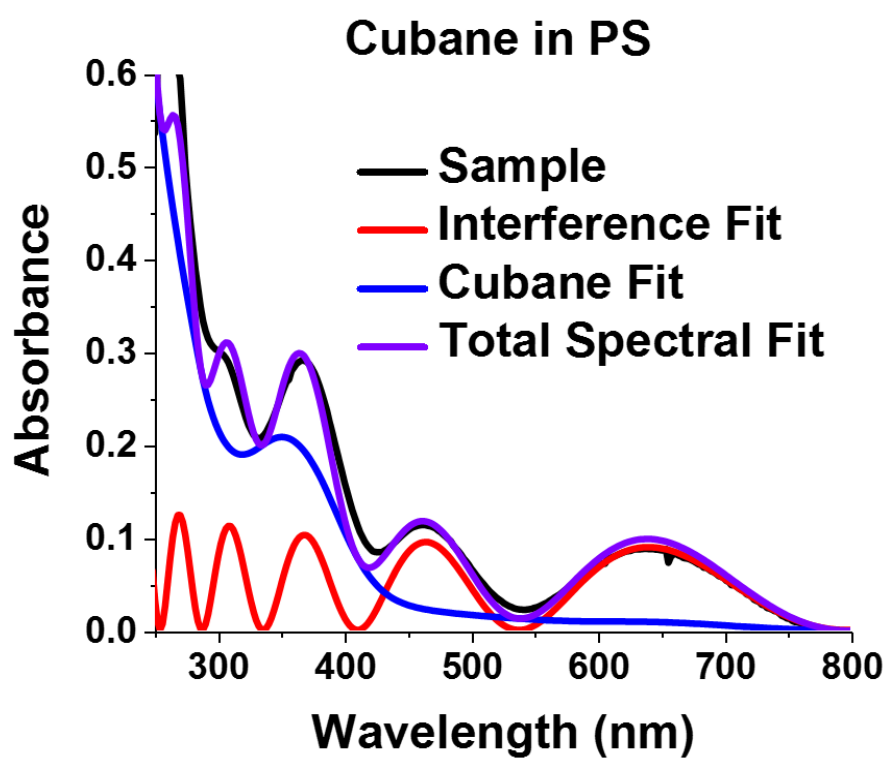

**Figure S2** UV-visible spectra of cubane embedded in PS and its spectral components.

**References**

Andersen, P., Andersen, T., Folkmann, F., Ivanov V, K., Kjeldsen, H. & West, J. B. (2001). *J. Phys. B At. Mol. Opt. Phys.* **34**, 2009–2019.

Chakrabarty, R., Bora, S. J. & Das, B. K. (2007). *Inorg. Chem.* **46**, 9450–9462.

Chiuzăian, S. G., Ghiringhelli, G., Dallera, C., Grioni, M., Amann, P., Wang, X., Braicovich, L. & Patthey, L. (2005). *Phys. Rev. Lett.* **95**, 197402.

Huibers, P. D. T. & Shah, D. O. (1997). *Langmuir*. **13**, 5995–5998.

Jaeger, F. M. & van Dijk, J. A. (1936). *Zeitschrift Fur Anorg. Und Allg. Chemie*. **227**, 273–327.

Jones, S. H., King, D. & Ward, A. D. (2013). **15**, 20735–20741.

Khusnutdinova, J. R., Luo, J., Rath, N. P. & Mirica, L. M. (2013). *Inorg. Chem.* **52**, 3920–3932.

Stafford, C. M., Roskov, K. E., Epps, T. H. & Fasolka, M. J. (2006). *Rev. Sci. Instrum.* **77**,.

Vura-Weis, J., Jiang, C. M., Liu, C., Gao, H., Lucas, J. M., De Groot, F. M. F., Yang, P., Alivisatos, A. P. & Leone, S. R. (2013). *J. Phys. Chem. Lett.* **4**, 3667–3671.

Wang, H., Young, A. T., Guo, J., Cramer, S. P., Friedrich, S., Braun, A. & Gu, W. (2013). *J. Synchrotron Radiat.* **20**, 614–619.